\documentclass[12pt]{revtex4}

\begin{document}
\title  {Distinguishability of complete and
unextendible product bases}
\author  { S. De Rinaldis$^{1,2,3,*}$}
\address{ $^1$NNL- National Nanotechnology Laboratory of INFM, via 
per Arnesano, 73100, Lecce, Italy;\\$^2$ISUFI- Istituto Superiore
Universitario per la Formazione Interdisciplinare, via per  
Arnesano, 73100, Lecce, Italy;\\$^3$IBM T.J. Watson Research Center, 
P.O. Box 218, Yorktown Heights, New York 10598, USA}
\date{\today}

\begin{abstract}

It is not always possible to distinguish multipartite orthogonal states if 
only local operation and classical communication (LOCC) are allowed. We prove 
that we cannot distinguish the states of an unextendible product bases 
(UPB)  by LOCC even with infinite resources (infinite-dimensional 
ancillas, infinite number of operations). Moreover we give a necessary and 
sufficient condition for the LOCC distinguishability of a complete 
product bases.

{$^*) Electronic$ mail: srinaldi@chem.utoronto.ca}\newline
Present address: Chemical Physics Theory Group, Department of 
Chemistry University of Toronto, 80 St. George Street, Toronto, Ontatio 
M5S 3H6, Canada\newline
\end{abstract}

\maketitle

In quantum mechanics orthogonal quantum states can always be 
distinguished. This is not always true when we restrict the set of actions 
on the multipartite system to LOCC only. 
About this topic a number of result has been proved: three Bell states
can never be distinguished \cite{Ghosh}, two orthogonal states can always
be distinguished \cite{Walgate1}, a characterization of the $2{\times}n$
states that can be distinguished by LOCC has been given \cite{Walgate2}, 
also conditions for LOCC unambiguos state discrimination have been derived 
\cite{chefles}. Surprisingly there are
pure orthogonal product vectors  that can be distinguished only
globally \cite{Bennett1}.  
More recently this effect has been further studied in \cite{mhorodecki}, 
with the result that set of states with less average entanglement 
than others (that are distinguishable by LOCC) can be undistinguishable.
Also the impact of restricted classical communication on 
distinguishability has been recently investigated \cite{hillery}. 
As an application of these concepts we mention a quantum communication 
protocol called  "data hiding" \cite{terhal}.\newline 
In this 
paper we prove that a class of product 
states, the unextendible
product bases (UPB), cannot be distinguished by LOCC, and give a necessary
and sufficient condition for the distinguishability of complete product  
bases. This fact proves that there is an entire class of separable 
superoperators that cannot be implemented by LOCC.
The part on UPB is different from the proof given in \cite{DiVincenzo},  
since there is no restriction to a finite number of rounds of communication 
and on the dimension of ancillary space exploited for performing 
generalized measurements (also in \cite{chefles} the results are 
restricted to "finite" resources).\newline
 \textbf {Definition 1.} We say that
we cannot distinguish "perfectly" a set of states by LOCC if we cannot
distinguish between them even using an infinite number of resources
(infinite number of LOCC "rounds", infinite dimensional ancillas, etc.)
while "exact" distinguishability is defined when finite resources are
used.\newline 
The distinction could appear of little importance if we think that in 
practical situations we never have an infinite amount or resources, but it 
seems significant if we restate it in terms of information. 
If we cannot distinguish exactly, but perfectly, between a set of states 
then we can acquire as much information as we want about the states, 
therefore we could optimize the amount of resorces employed versus 
the information attainable. If the states cannot be distinguished 
perfectly, 
then 
the information we can obtain between them is upperbounded by a 
finite amount. In terms of  superoperators theory this implies 
that we have found an entire class of separable superoperators 
that are not in the class of LOCC superoperators\cite{Rains}.\newline 
\textbf {Definition 2.} Consider a multipartite Hilbert space 
$H=H_1{\otimes}H_2{\otimes}...{\otimes}H_n$ and a product bases that 
span a space $H_{PB}$. 
An unextendible product bases (UPB) \cite{DiVincenzo} is a product bases 
for which the 
complementary subspace $H^{\perp}_{PB}$ does not contain product 
vectors.\newline
Let us introduce the concept of
"irreducible UPB".\newline 
\textbf {Definition 3.}
An "irreducible UPB" is an unextendible product bases 
in $H_A{\otimes}H_B$
that
cannot be divided in two set of vectors contained in the subspaces
$H'_A{\otimes}H_B$ and
$H'^{\perp}_A{\otimes}H_B$ or $H_A{\otimes}H'_B$ and
$H_A{\otimes}H'^{\perp}_B$.\newline
Every UPB contains an "irreducible UPB" in one of its subspaces. It is
trivial
to prove that if this were not the case than the UPB would be a complete
product bases.
UPB have been studied for their properties related to bound entanglement 
\cite{horodecki}.
Bennett et al. \cite{Bennett1} have shown a set of nine orthogonal 
product
states that
cannot be perfectly distinguished by LOCC. 
This is the only example known to us. Are there other product 
states that are not perfectly distinguishable? 
In this paper we answer to this question by showing a class of product 
states, the UPB,  that can never be 
perfectly distinguished by LOCC. It has already been proven that UPB 
cannot be 
exactly 
distinguishable \cite{DiVincenzo2}. This is relevant because it proves 
that there is an entire class of separable superoperators that cannot be 
implemented by LOCC, i.e. the two classes are {\it not} equal except a 
few particular cases.

\textbf {Theorem 1.} 
We cannot perfectly distinguish an UPB (unextendible product bases) by 
LOCC operations.\newline
\textbf {Proof}.
Let us consider first a bipartite UPB: 
${\{}|{\psi_i}{\rangle}=|{\phi_i}{\rangle}|{\chi_i}{\rangle}{\}}$. 
We will prove that the effect on every state of a POVM element we can 
apply, without 
creating 
nonorthogonal states, is either to eliminate a state or to create a state 
parallel 
to the previous one. 
Let us consider an Alice  POVM element $E$. It is an hermitian operator, 
so 
it is 
diagonal in an orthonormal bases 
$|0{\rangle}{\langle}0|,...,|N{\rangle}{\langle}N|$.
We expand the set of vectors ${\{}|{\phi_i}{\rangle}{\}}$ in this bases:

\begin{eqnarray}
\nonumber|{\psi_0}{\rangle}=|0{\rangle}c_{00}|{\chi_0}{\rangle}+
|1{\rangle}c_{10}|{\chi_0}{\rangle}+{\qquad}
{\cdot}{\qquad}+|N{\rangle}c_{N0}|{\chi_0}{\rangle}\\
\nonumber{\qquad}{\cdot}{\qquad}{\qquad}{\cdot}{\qquad}{\qquad}
{\cdot}{\qquad}{\qquad}{\cdot}{\qquad}{\qquad}{\qquad}{\cdot}{\qquad}\\
\nonumber|{\psi_l}{\rangle}=|0{\rangle}c_{0l}|{\chi_l}{\rangle}+
|1{\rangle}c_{1l}|{\chi_l}{\rangle}+{\qquad}
{\cdot}{\qquad}+|N{\rangle}c_{Nl}|{\chi_l}{\rangle}\\
\nonumber{\qquad}{\cdot}{\qquad}{\qquad}{\cdot}{\qquad}{\qquad}
{\cdot}{\qquad}{\qquad}{\cdot}{\qquad}{\qquad}{\qquad}{\cdot}{\qquad}\\
|{\psi_k}{\rangle}=|0{\rangle}c_{0k}|{\chi_k}{\rangle}+
|1{\rangle}c_{1k}|{\chi_k}{\rangle}+{\qquad}
{\cdot}{\qquad}+|N{\rangle}c_{Nk}|{\chi_k}{\rangle}  
\end{eqnarray}

Let us suppose that $E$ is nonzero on $|{\phi_0}{\rangle}$. 
Since the 
resulting vectors 
${\{}E{\otimes}I|{\psi_i}{\rangle}=(E|{\phi_i}{\rangle})|{\chi_i}{\rangle}{\}}$
must remain orthogonal, the vectors orthogonal to $|{\phi_0}{\rangle}$ 
must 
remain orthogonal after the application of $E$, that is 
${\langle}{\phi_i}|{\phi_0}{\rangle}=0{\Longrightarrow{\langle}
{\phi_i}|E^{\dagger}E|{\phi_0}{\rangle}=0}$.
We write E in the diagonal bases: 
$E={\lambda_0}|0{\rangle}{\langle}0|+...+{\lambda_N}|N{\rangle}{\langle}N|$,
where the ${\{}{\lambda_i}{\}}$ are real positive numbers less than one.

The orthogonality condition translates into the equations:

\begin{equation}
c^*_{0i}{\lambda^2_0}c_{00}+...+c^*_{Ni}{\lambda^2_N}c_{N0}=0
\end{equation}

for all the vectors for which : 

\begin{equation}
c^*_{0i}c_{00}+...+c^*_{Ni}c_{N0}=0.
\end{equation}

The condition above means that the product vector 
$|{\psi'_0}{\rangle}=|0{\rangle}{\lambda^2_0}c_{00}|{\chi_0}{\rangle}+
|1{\rangle}{\lambda^2_1}c_{10}|{\chi_0}{\rangle}+
......+|N{\rangle}{\lambda^2_N}c_{N0}|{\chi_0}{\rangle}$ is 
orthogonal to all the vectors to which $|{\psi_0}{\rangle}$ is orthogonal.
The vector $|{\psi'_0}{\rangle}$ must be parallel to $|{\psi_0}{\rangle}$, 
because if not we could construct the vector  $|{\psi'_0}{\rangle}- 
{\langle}{\psi_0}|{\psi'_0}{\rangle}|{\psi_0}{\rangle}$
that is orthogonal to all the vectors of the UPB, thus against the 
assumption that 
the product bases is unextendible . 
Even if until now we have considered only local measurement, i.e. we have 
restricted 
the set of Alice operators to  POVM elements, our results holds also in 
the 
general case. In fact, Alice action is described by a 
superoperator and for every 
operation element  S, from the polar decomposition theorem, S is a 
product of a unitary (U) and a positive (E) operator: S=EU (right polar 
decomposition). 
We have
$S|{\phi_i}{\rangle}=(EU|{\phi_i}{\rangle}=E|{\phi'_i}{\rangle}$
where the set ${\{}|{\phi'_i}{\rangle}{\}}$ is an UPB because an UPB is 
tranformed in another UPB with a unitary operation U. It is trivial to see 
that if we could extend the bases to  a new orthogonal product vector then 
we could apply 
$U^{-1}$ to this vector to obtain a new product vector orthogonal to the 
previous set, unextendible for assumption. Therefore there is no loss of 
generality in considering only 
local measurement. 
The new set of vector ${\{}E|{\psi_i}{\rangle}{\}}$ is an UPB in the 
subspace spanned by the 
vectors that constitute the base in which E is diagonal. In fact if we 
could extend the 
product bases in this subspace to 
another  product vector, this vector would be orthogonal also to the ones 
eliminated by E and therefore the starting base would be extendible.
In general the set ${\{}E|{\psi_i}{\rangle}{\}}$ could be a 
complete bases that, by definition, is a "trivial" UPB because it 
also has the 
property that we cannot find another product
state orthogonal to all the member of the bases. However, in a local 
measurement with POVM elements ${\{}E_l{\}}$, since for what we have 
proved, the operators 
$E_l$ 
are either orthogonal or proportional, not all the sets 
${\{}E_l|{\psi_i}{\rangle}{\}}$ can be 
complete bases unless the starting set ${\{}|{\psi_i}{\rangle}{\}}$ is a 
complete 
base. From the property of the set ${\{}E_l{\}}$, we notice that even if 
we have 
an 
infinite number of elements in the set, only  a finite number of outcomes 
are different. 
To prove the theorem excluding that we could distinguish with an infinite 
number of rounds we notice that, since the only two operations that we can 
perform with a measurement on a state is either to leave the state 
unchanged or to 
eliminate it, if we want that they remain orthogonal,
at some point,
when we could not eliminate  other
states, the only POVM that we could apply is proportional to the identity.
However it is not sufficient to show that at some point of the LOCC 
protocol the state must become nonorthogonal, because in principle an 
infinite set of weak measurement strategies \cite{weak}
is possible and if 
the states 
at every protocol step are "nearly" orthogonal they could still be 
distinguished. This is completely general, as proved by construction in 
\cite{Bennett1}, because any strategy involving weak and strong 
measurement 
can be replaced by a strategy involving only weak measurement. 
To complete the proof we must show that at some point if we want to 
acquire information about the states they should become nonorthogonal by a 
finite amount. At this point we will show that the mutual information 
between the measurement outcome and the state is less than the information 
obtainable by a nonlocal measurement.  
We will restrict the attention to an "irreducible UPB" and prove that the 
information attainable about the state of an 
irreducible 
UPB is upperbounded by $O({\delta})$ where ${\delta}$ is the maximum 
overlap between two vectors of the new set of states.
Since every UPB contains an "irreducible UPB" then it will follow that 
also the set of states forming the UPB are not distinguishable by LOCC.
Let us consider an irreducible UPB and the first Alice operation. If we 
want that the states remain orthogonal only an operator proportional to 
the identity 
is possible. In fact since we have proved that a POVM element either 
eliminate a vector or leave it unchanged, then we could either eliminate 
eliminate some vector or leave all unchanged.
The first case leads to a contradiction because we could divide the set of 
states of the UPB in two sets: the vectors eliminated  in 
$H'_A{\otimes}H_B$ and the others in
$H'^{\perp}_A{\otimes}H_B$, in constrast to the definition of irreducible 
UPB. If we want to leave all the vector unchanged then we must apply 
an operator proportional to
the identity.
Therefore if we want that the states are 
"nearly" orthogonal we must use an 
operator of the form $E={\lambda}I+{\lambda}{\delta'}A$,
where  ${\lambda}$ is a real positive number less than one,  
${\delta'}$ is an infinitesimal real positive number related to the 
maximum overlap among the new set of vectors and A is a positive operator.
The maximum overlap between two states is:

\begin{equation}
max_{i,j} {\langle} 
{\phi_i}|E^{\dagger}E|{\phi_j}{\rangle}= 2{\delta'}{\langle}
{\phi_i}|A|{\phi_j}{\rangle}+{\delta'^2}
{\langle}{\phi_i}|A^{\dagger}A|{\phi_j}{\rangle}>2{\lambda}^2{\delta'}{\langle}
{\phi_i}|A|{\phi_j}
{\rangle}={\delta'}c
\end{equation}

where c is a real number.
We define $p({\phi_i},m)$ as the probability that, once obtained the 
measurement result m, the state is $|{\phi_i}{\rangle}$. The probabilities 
before starting the protocol are all the same. We define:

\begin{equation}
{\epsilon} =max_i p({\phi_i},m)- \frac{1}{n}
\end{equation}

where ${\epsilon}$ is the maximum amount of information we can obtain 
about a state.

From the definition we have:

 \begin{equation}
p({\phi_i},m) =
\frac{{\langle}
{\phi_i}|E^{\dagger}_mE_m|{\phi_i}{\rangle}}{\sum_j{{\langle}
{\phi_j}|E^{\dagger}_mE_m|{\phi_j}{\rangle}}} 
\end{equation}

If we define $a_j=2{\langle}{\phi_j}|A|{\phi_j}{\rangle}$
we have, neglecting the terms in ${\delta'}^2$ :

 \begin{equation}
p({\phi_i},m) = \frac{1+{\delta'}a_i}{n+{\delta'}{\sum}_j a_j}
{\thinspace}{\leq}{\thinspace} \frac{1}{n}+ \frac{{\delta'}{\sum}_j 
a_j}{n^2}+ 
\frac{{\delta'}a_i}{n} 
\end{equation}

Therefore

 \begin{equation}
{\epsilon} = p({\phi_i},m) -  \frac{1}{n}
{\thinspace}{\leq}{\thinspace} {\delta'}( \frac{{\sum} a_j}{n^2}+ 
\frac{a_i}{n})
\end{equation}    

This last equation means that if we want to acquire a finite amount of 
information then also 
the states are nonorthogonal by a finite amount.
Let us consider N rounds of measurement. We can write a general operation 
element 
implemented by LOCC as \cite{note0} :

 \begin{equation}
S_m = A_m {\otimes} B_m 
\end{equation}

 \begin{equation}
A_m = E_NE_{N-1}..E_1
\end{equation}

 \begin{equation}
B_m = F_NF_{N-1}..F_1 
\end{equation} 

where $E_i$ and $F_i$ are positive operators.
We can consider only product of positive operators. In fact let us 
consider a general separable operator 
$S'_m=A'_m {\otimes} B'_m$ with $A'_m=H_NH_{N-1}..H_1$ and 
$B'_m=K_NK_{N-1}..K_1$. 
We can construct an operator $S_m=A_m {\otimes} B_m$ 
(with positive operators, notation as in equations 9, 10, 11) such that 
${\langle}{\phi_i}|S'^{\dagger}_mS'_m|{\phi_i}{\rangle}=
{\langle}{\phi_i}|S^{\dagger}_mS_m|{\phi_i}{\rangle}$.
We use first a left polar decomposition : $H_i=U_iE'_i$ and we have:
$H_m=U_NE'_NU_{N-1}E'_{N-1}..U_1E_1$, then we take all the unitary 
operators 
to the left, thanks to the fact that every linear operator has a left 
and a right polar decomposition: $E'_1U_1=U_2E'_2$. After some steps we 
arrive at a 
"generalized" polar decomposition: $A'_m=U_NU_{N-1}..U_1E_NE_{N-1}..E_1$ 
(similarly for $B'_m$). 
Therefore the result is formally equivalent to a product of positive 
operators.

To maintain the states nearly orthogonal in every round we must have:  
$E_i={\lambda}_iI+{\lambda}_i{\delta'}A_i$ and 
$F_i={\rho}_iI+{\rho}_i{\delta'}B_i$. Following the same procedure of the 
single step case we have that the 
overlap between two states is (neglecting the terms superior 
to first order in  ${\delta'}$):

    \begin{eqnarray}
\nonumber{\thinspace}{\delta}= max_{j,k} {\delta}_{jk}=
max_{j,k} {\langle}
{\phi_j}|S^{\dagger}S|{\phi_k}{\rangle}={\qquad}{\qquad}\\
{\thinspace}max_{j,k} 
{\sum}_i(2{\delta'}{\langle}
{\phi_j}|A_i{\otimes}I|{\phi_k}{\rangle}+
2{\delta'}{\langle}{\phi_j}|I{\otimes}B_i|{\phi_k}{\rangle})=
max_{j,k} {\delta'}{\sum}_i(a_{ijk}+b_{ijk})
\end{eqnarray}

where $a_{ijk}=2{\langle}{\phi_j}|A_i{\otimes}I|{\phi_k}{\rangle}$ and  
$b_{ijk}=2{\langle}{\phi_j}|I{\otimes}B_i|{\phi_k}{\rangle}$

Following the same calculations that lead to equation (8) we can find 
that:

 \begin{equation}
{\epsilon} = p({\phi_j},m) -  \frac{1}{n}
{\thinspace}{\leq}{\thinspace}{\delta'}{\sum}_i(c_{ij}+d_{ij}) 
\end{equation}

where $c_{ij}= \frac{{\sum}_j a_{ij}}{n^2}+ \frac{ 
a_{ij}}{n}
$ and $d_{ij}= \frac{{\sum}_j b_{ij}}{n^2}+ \frac{b_{ij}}{n}
$ ($a_{ij}=2{\langle}{\phi_j}|A_i{\otimes}I|{\phi_j}{\rangle}$ and 
$b_{ij}=2{\langle}{\phi_j}|I{\otimes}B_i|{\phi_j}{\rangle}$).

In order to find a relation analog to equation (4) we notice that formally 
we are in the same situation but with the operator 
$O(N)={\sum_{i=1}^N} A_i{\otimes}I+I{\otimes}B_i$
and we find, analog to (8):

  \begin{equation}
{\epsilon}_N {\thinspace}{\leq}{\thinspace}{\delta'}( \frac{{\sum} 
a_j}{n^2}+ 
\frac{a_i}{n})={\delta'}M_N
\end{equation}

where $a_j={\langle}{\phi_j}|O(N)|{\phi_j}{\rangle}$
and :

\begin{equation}
max_{j,k} {\langle}
{\phi_j}|S^{\dagger}S|{\phi_k}{\rangle}={\delta}={\delta'}c_N
\end{equation}

where $c_N=max_{j,k} {\langle}{\phi_j}|O(N)|{\phi_k}{\rangle}$.
We arrive a the final expression:

  \begin{equation}
{\epsilon}_N {\thinspace}{\leq}{\thinspace}{\delta}\frac{M_N}{c_N}
\end{equation}

Let us consider the behaviour of O(N) when $N{\rightarrow}{\infty}$.
We examine the different cases. If $||O(N)||{\rightarrow}{\infty}$ we can  
write $O(N)=K_NO(N)$
where $K_N{\rightarrow}{\infty}$ and $||O'(N)||{\rightarrow}a$ a real 
number, so the ratio $\frac{M_N}{c_N}$ is finite because the 
$K_N$ in the ratio cancel. The same argument holds if 
$||O(N)||{\rightarrow}0$. 
If O(N) tends to 
a multiple of the identity when $N{\rightarrow}{\infty}$ then 
$c_N{\rightarrow}0$ but not $M_N$, so we cannot bound ${\epsilon}$ 
with a multiple of ${\delta}$ as in (16).
However we can easily see that in this case we 
do not need the bound (16) because it is easy to see that 
we cannot extract a finite amount of 
information about the states. In fact from (5) 
and (6) we can easily calculate that 
${\epsilon}{\rightarrow}0$ \cite{note1} .
We conclude that if we mantain the states nonorthogonal by an 
infinitesimal 
amount we cannot reach a finite amount of information about them.
The generalization to N-parties states i straightforward. It simply leads 
to a redefinition of O(N); for example for three parties it becomes:
$O(N)={\sum_{i=1}^N} A_i{\otimes}I{\otimes}I+I{\otimes}B_i{\otimes}I+
I{\otimes}I{\otimes}C_i$ and the conclusions are the same.

Now let us consider the case in which the state are nonorthogonal by a 
finite amount ${\delta}$ at Nth measurement round, that we consider stage 
I. The stage II is when the protocol is completed. We will generalize the 
argument in \cite{Bennett1}, that, indeed, is very general, i.e. do not 
depends neither on the 
number of parties nor on the number of  states, finding a bound for 
the mutual information attainable.  
We use the same notation of \cite{Bennett1};  $M_I$ ($M_{II}$) is the 
random 
variable 
describing the stage-I (stage-II) outcomes; W is the variable that figures 
out which of the states has been measured; $I(W;M_I,M_{II})$ is the 
mutual information between the measurement outcomes $M_I$, $M_{II}$ and 
W. Using the additivity property and the definition of mutual information 
we find:

\begin{equation}
I(W;M_I,M_{II})=log_2n - {\sum}_{m_I} p(m_I)[H(W|m_I)-I(W;M_{II}|m_I)]
\end{equation}
 
where n is the number of states to be distinguished, $p(m_I)$ is the 
probability of outcome $m_I$ of the measurement in stage I, H is the 
entropy function. 
At the end of stage I the states are 
${\rho}_i=|{\phi}_{i,m_I}{\rangle}{\langle}{\phi}_{i,m_I}|$ with 
probabilities $q_i=p({\psi}_i|m_I)$ and ${\{}M_b{\}}$ is a positive operator 
valued measure performed in stage II. Let us consider the two states that 
are 
nonorthogonal at stage I 
${\langle}{\phi}_{1,m_I}|{\phi}_{2,m_I}{\rangle}={\delta}$ and divide the 
density operator in two part:

\begin{equation}
{\tau}_1={\sum_{i=1}^2}\frac{q_i}{s_1}{\rho}_i,{\qquad}{\qquad}{\tau}_2=
{\sum_{i=3}^n}\frac{q_i}{s_2}{\rho}_i
\end{equation}

with $s_1=q_1+q_2$ and $s_2=1-s_1$. We have 
${\rho}=s_1{\tau}_1+s_2{\tau}_2$.
Using the concavity of Shannon entropy and removing the dependence of all 
the states except the first two we arrive at the expression:

  \begin{eqnarray}
\nonumber{\thinspace}H(W|m_I)-I(W;M_{II}|m_I){\thinspace}{\geq}
{\thinspace}2[(\frac{1}{n}-(n-1){\epsilon})]{\cdot}{\qquad}{\qquad}\\
{\thinspace}[1+{\sum}_b (tr{\tau}_1 M_b) log_2 (tr{\tau}_1M_b)- 
{\sum_{i=1}^2}
\frac{1}{2} 
{\sum}_b(tr{\rho}_iM_b)log_2 (tr{\rho}_iM_b)]
  \end{eqnarray}

Minimizing the expression above as in \cite{Bennett1} we 
find:

\begin{equation}
H(W|m_I)-I(W;M_{II}|m_I){\thinspace}{\geq}{\thinspace} 2[( \frac{1}{n} - 
(n-1){\epsilon}) h( \frac{1}{2}- \frac{1}{2} {\sqrt{1-{\delta}^2}})]
\end{equation}

The quantity in (20) is strictly positive if ${\delta}>0$.

Therefore we conclude that $I(W;M_I,M_{II})<log_2n$ if the states at 
some 
stage of the protocol are nonorthogonal by a finite amount. 
Note that the part (iii) of the proof is valid for a general set of 
states and measurements. The 
extension to the multipartite case is immediate. This 
completes the proof.\newline
\textbf {Theorem 2} \cite{note2}. \newline
A complete product bases is distinguishable by LOCC {\it if and only 
if} it 
does not contain an "irreducible UPB".
Moreover, if a complete product bases is 
distinguishable by LOCC, then it is
distinguishable by a protocol consisting only in von Neumann 
measurements.\newline
\textbf {Proof.} 
The proof follows from the results on UPB; in fact a complete bases  is a 
trivial UPB because it has the property that we cannot find another product 
state orthogonal to all the member of the bases.
Therefore, as we have proven for Theorem 1, if the complete bases contains 
an irreducible UPB, then the information attainable about that set of 
states is less, by a finite amount, then the maximum information.
If a complete product bases does not contain an "irreducible UPB", by 
definition, we can divide the states in two set of vectors contained 
in the subspaces
$H'_A{\otimes}H_B$ and
$H'^{\perp}_A{\otimes}H_B$ or $H_A{\otimes}H'_B$ and
$H_A{\otimes}H'^{\perp}_B$. This fact gives a procedure for distinguishing 
the states by a protocol consisting only in von Neumann measurements: we 
use the projectors 
$P_A$ and $P^{\perp}_A$ (or $P_B$ and $P^{\perp}_B$) that project, 
respectively, on subspace $H'_A{\otimes}H_B$ and
$H'^{\perp}_A{\otimes}H_B$ (or $H_A{\otimes}H'_B$ and
$H_A{\otimes}H'^{\perp}_B$). We can iterate this procedure until only one 
state remains, so we have successfully completed the task. 
This completes the proof.\newline
\textbf {Remark.}
Since it can be not always obvious to check if a complete bases 
contains or not an "irreducible UPB", we can give a method to check 
the perfect distinguishability 
of
a complete bases with a simple algorithm, without involving lenght
calculations.
The method works as follows:
let us first consider the Alice vector and construct an ensemble; we start 
with one  vector and 
find all the vectors that are nonorthogonal to it; we have now a set of 
vectors; we expand this set performing a series of steps in each one we 
find the vectors 
nonorthogonal to at 
least one member of the set. Since a POVM element that is nonzero on one 
vector of this set must have as eigenvectors all the vectors of the set 
for construction, then it could be only the identity in the subspace 
spanned 
by the vectors of the set. Thus if this protocol finds all 
the vectors of the bases, then the only POVM element we can apply is the 
identity. If the same holds also for Bob vectors, then whatever POVM 
elements we apply (except the identity) we create nonorthogonal states and 
therefore we cannot perfectly distinguish the states.
In general if we find only a subset of the total set of vectors, with a 
von Neumann measurement we split in two the total set of states.
After that the protocol continues with classical communication to Bob;
now Bob repeats the same procedure. This protocol continues until either 
we distinguish the states or we arrive at a point where only the identity 
can be applied (that means that we have found an "irreducible UPB").

Note that at 
most ${\sum}n_j$ steps (therefore  $({\sum}n_j)-1$ bits of classical 
communication), where $n_j$ are the dimensions of the 
multipartite 
Hilbert space, are necessary to 
distinguish 
between the states, since every step must eliminate at least one 
dimension of the total space. Therefore the number of bits grows 
at most linearly, whereas the number of 
states grows exponentially with the number of parties.\newline 
\textbf {Example}.
As a corollary of Theorem 2 we can answer to the question (posed in 
\cite{Bennett1}) of LOCC
distinguishability of the Lagarias-Shor 1024 state ten-parties
complete bases \cite{Shor}.
Every party has a qubit which is one state 
out of
$|0{\rangle}$, $|1{\rangle}$, $|0+1{\rangle}$, $|0-1{\rangle}$.
Since for every party in the set of 1024 states there are all the
four states above, then the states cannot be divided in two
orthogonal subspaces, therefore this complete bases is an irreducible UPB.
We conclude that  this bases is not perfectly
distinguishable by LOCC.

\section {Acknowledgements}
The main part of this work was completed at IBM T.J. Watson research 
center. I would like to thank the Quantum information group at IBM for 
their
hospitality, Dr. David DiVincenzo for useful discussion, advice and 
careful reading of the manuscript, Dr. Charles Bennett and Dr. Barbara 
Terhal for 
helpful discussions.


\begin{thebibliography}{aaaaaaaaaaaaaaaaaaaa}
\bibitem{note0}This is only a necessary condition for writing a LOCC 
operator. Actually in the proof we never use neither that in every 
round the POVM element must sum to the identity nor classical 
communication. In fact the Part II of the proof  is valid for separable 
operators, but only for the ones "near" the identity, since a complete 
product bases is always distinguishable by separable superoperators.
\bibitem{note1} The fact, proved in part I, that there are not POVM 
elements (except multiples of the identity) 
of the form $A{\otimes}I$ or $I{\otimes}B$ (note that it is also true for 
operators of the form  $A{\otimes}I+I{\otimes}B$ as O(N)) 
that keeps the states orthogonal, is essential. If not we could have  
${\epsilon}>0$ and ${\delta}=0$ after stage I.
\bibitem{note2}In the particular case of the nine domino-like state of 
\cite{Bennett1}, in \cite{Walgate2} 
and \cite{Vaidman} a similar argument is applied showing that a POVM 
different from the unitary operator leads to nonorthogonality of states. 
However 
it 
requires to write  
the equations for the specific case and to prove that they have 
no solutions.
\end{thebibliography}
\end{document}